\documentclass[twocolumn,prl,showpacs,superscriptaddress,reprint]{revtex4-1}
\usepackage[resetlabels]{multibib}
\newcites{sec}{Supplementary References}

\usepackage{graphicx}

\usepackage{amsthm,color,amsfonts,graphicx,verbatim}
\usepackage{amsmath}

\usepackage[normalem]{ulem}

\usepackage{amsfonts}
\usepackage{listings}
\usepackage{enumerate}
\usepackage{latexsym}
\usepackage{bm}
\usepackage[breaklinks,colorlinks = true,linkcolor = red,urlcolor=cyan,citecolor=red]{hyperref}
\newcommand{\beq}{\begin{equation}}
\newcommand{\eeq}{\end{equation}}
\newcommand{\be}{\begin{eqnarray}}
\newcommand{\ee}{\end{eqnarray}}

\newcommand{\bra}[1]{\left\langle{#1}\right|}
\newcommand{\ket}[1]{\left|{#1}\right\rangle}

\renewcommand{\paragraph}[1]{\vspace{4pt}\noindent {\it #1.---}}

\begin{document}
\title{Non-Fermi glasses: fractionalizing electrons at finite energy density}
\author{S. A. Parameswaran}
\affiliation{Department of Physics and Astronomy, University of California, Irvine, CA 92697, USA}

\author{S. Gopalakrishnan}
\affiliation{Department of Physics and Burke Institute, California Institute of Technology, Pasadena, CA 91125, USA}

\date{\today}
\begin{abstract}
Non-Fermi liquids are metals that cannot be adiabatically deformed into free fermion states. We argue for the existence of ``non-Fermi glasses,'' which are phases of interacting disordered fermions that are fully many-body localized, yet cannot be deformed into an Anderson insulator without an eigenstate phase transition. We explore the properties of such non-Fermi glasses, focusing on a specific solvable example. At high temperature, non-Fermi glasses have qualitatively similar spectral features to Anderson insulators. We identify a diagnostic, based on ratios of correlation functions, that sharply distinguishes between the two phases even at infinite temperature. We argue that our results and diagnostic should generically apply to the high-temperature behavior of the many-body localized descendants of fractionalized phases.
{}
\end{abstract}
\maketitle
Adiabatic continuity, exemplified by the Fermi liquid, is a central theme in many-body physics~\cite{pwabook}. The interactions in Fermi liquids only ``dress'' the elementary excitations, which retain the character of the microscopic fermions; the quasiparticle residue $Z$, which measures the overlap between the quasiparticle and bare fermion creation operators, remains finite, and the (bare) electron spectral function---measured by tunneling experiments---exhibits sharply dispersing quasiparticle modes. In the presence of strong correlations, however, the Fermi liquid picture can break down, as $Z \rightarrow 0$ at a quantum phase transition to a ``non-Fermi liquid'' phase. Non-Fermi liquid phases (in common with other exotic states such as spin liquids) often exhibit fractionalization: their elementary excitations have distinct quantum numbers from the bare fermions.
 For instance, it is possible for an electron to be fractionalized into elementary collective excitations that carry spin and charge, and have widely separated velocities. The single-electron spectral function will then typically not exhibit sharply dispersing modes, but rather an incoherent, multi-particle continuum. Such distinctions might not, however, manifest in transport: for example, in the simplest type of non-Fermi liquid, the ``orthogonal metal''~\cite{OrthogonalMetals}, the transport and thermodynamic properties are \emph{identical} to those of a Fermi liquid, but the single-particle spectral properties are dramatically different. 

In clean systems such as Fermi liquids, adiabatic continuity only applies to elementary excitations above the ground state; however, recent~\cite{BAA,husereview} work on strongly disordered systems in the many-body localized (MBL) phase indicates that in such systems, the entire many-body spectrum might exhibit a form of adiabatic continuity. In particular, a typical many-body eigenstate of an MBL system can be regarded (almost everywhere~\cite{jzi}) as a product state (or Slater determinant) of localized orbitals, perturbatively dressed by interactions~\cite{PhysRevB.90.174202}, much as the Fermi liquid is viewed as a dressed version of a Fermi gas --- an apt name might be a ``Fermi glass''~\cite{AndersonFermiGlass,Fermiglass}. This leads us naturally to ask: do there exist fractionalized, ``non-Fermi glass'' MBL phases, which cannot be regarded as perturbatively dressed Slater determinants; and if so, what are their properties?

In this work, we 
embark on the study of non-Fermi glasses by exploring the fate of the ``orthogonal metal''~\cite{OrthogonalMetals} in the presence of strong quenched disorder. We term the resulting phase the orthogonal many-body localized insulator. Although the oMBL insulator has the same transport and thermodynamic properties as the Fermi glass, or conventional MBL (cMBL) insulator, the two phases are separated by an eigenstate phase transition~\cite{HuseMBLQuantumOrder,BauerNayak}.
 We discuss the properties of the oMBL phase using an exactly solvable one-dimensional toy model as a concrete example (though our conclusions apply more generally). 

The usual spectral signatures of fractionalization are absent in this strongly disordered limit. Since both the oMBL and cMBL phases are localized, the local single-particle spectral function in both phases is dominated by sharp peaks. {These are related to localized integrals of motion (LIOMs)~\cite{PhysRevB.90.174202}, that can be accessed by electron tunneling in both phases}. The \emph{spatially averaged} spectral functions retain a sharp distinction at zero temperature, featuring a soft gap in the oMBL phase but not in the cMBL phase; however, at nonzero temperatures this soft gap is filled in. Thus the eigenstate phase transition between oMBL and cMBL phases is spectrally ``hidden.'' Nonetheless, the two phases are distinct even at infinite temperature; this distinction is hidden in the structure of spatial correlations. We identify a specific ratio of correlation functions (namely, that of  the single-particle propagator to the pair propagator) that sharply distinguishes them. Such ratios of spatial correlators should generically diagnose  fractionalization in the MBL setting. 

\emph{Disordered orthogonal metals}.---We first schematically review the construction of the orthogonal metal~\cite{OrthogonalMetals},  
of spinless electrons. The electron operator at site $i$, $c_i$, is written as the product of a fermion operator $f_i$ and a Pauli matrix $\sigma^x_i$ acting on an auxiliary subspace. This doubling of the degrees of freedom on each site is associated with a $Z_2$ gauge redundancy~\cite{wenbook}. Thus, the original theory (with only `physical' $c$ fermions) can be recast as a theory of $f$ fermions and $\sigma$ spins (in this construction, subject to a transverse-field Ising-model (TFIM) Hamiltonian, $H_{\text{TFIM}} = h \sum_i \sigma^z_i + J \sum_{i} \sigma^x_i \sigma^x_{i+1}$) coupled to a fluctuating $Z_2$ gauge field. This has two possible $T\rightarrow0$ phases. When the $Z_2$ gauge theory is confining or the TFIM is in its ordered phase, the propagating degrees of freedom are the $c$ fermions, and the phase is a conventional metal. However, when the $Z_2$ gauge theory is deconfined and the TFIM is in its paramagnetic phase, the $c$ fermion is fractionalized into separately propagating $f$ and $\sigma$ degrees of freedom: the orthogonal metal phase. In spatial dimension $d>1$, where $Z_2$ gauge theory has both confined and deconfined phases, both orthogonal and conventional metals are possible. 

This phase structure readily generalizes to excited eigenstate properties in the presence of strong randomness. In two dimensions, pure $Z_2$ gauge theory with random couplings is related by duality to a random transverse-field Ising model (RTFIM). For strong randomness, the gauge theory has confined and deconfined ``eigenstate phases'' {that survive the inclusion of localized dynamical matter}~\cite{HuseMBLQuantumOrder}. The oMBL phase then exists in the regime where (1) the gauge theory is in its deconfined, localized eigenstate phase, (2) the $\sigma$ spins are in the localized paramagnetic phase, and (3) the $f$ fermions are localized~\footnote{We neglect for now the question of whether MBL in $d=2$ is destroyed by rare-region effects. Even if this is so, the phenomena we discuss will be visible as sharp crossovers.}. In this regime, the ``matter'' degrees of freedom ($f$s and $\sigma$s) are weakly coupled to the gauge sector, which can therefore be neglected. Note that at high temperatures, oMBL and cMBL phases are not the only possibilities; there is also a \emph{thermal} phase. If any of the three sectors is in its thermal phase, it will infect the other sectors~\cite{SpinCatalysis}; to avoid this issue, we focus on the strong randomness limit where all sectors are localized.

When the gauge field is deconfined, we can regard $f$ and $\sigma$ as separately propagating degrees of freedom, and the $c$ fermion as a composite of them, with the proviso that only gauge-neutral quantities (i.e., the $c$ fermions) are measurable. To discuss diagnostics of this regime in a simple solvable model, and to make contact with extant literature on MBL, we consider a $d=1$ toy model of interacting fermions and spins, on which we \emph{impose} the additional condition that only the {$c$} fermions are measurable. In $d=1$, the gauge theory is always confining, so such a model cannot arise by fractionalizing spinless fermions; however, the spectral and correlation properties of the $d=1$ model generalize directly to $d>1$.  

{\it Exactly Solvable Model.---} Our  $d=1$ model 
(introduced, absent disorder, in ~\cite{OrthogonalMetals}) consists 
of  fermions ($c_i$) and Ising spins ($\sigma_i$), with Hamiltonian 
\be\label{HES}
H_{es} & = & \sum\nolimits_i J_i \sigma^x_i \sigma^x_{i+1} + h_i \sigma^z_i (-1)^{c^\dagger_i c_i} \nonumber \\
& & \quad - t_i (c^\dagger_i \sigma^x_i \sigma^x_{i+1} c_{i+1} +\text{h.c.}) + \mu_i c^\dagger_i c_i
\ee
where $J_i$, $h_i$, $t_i$, and $\mu_i$ are random variables.
We now rewrite \eqref{HES} in terms of  $\tau^z_i \equiv \sigma^z_i (-1)^{f^\dagger_i f_i}$, $\tau^x_i = \sigma^x_i$, $f_i \equiv \tau^x_i c_i$: the two sectors decouple, $H_{es}  =  H_f + H_\tau$, with
\be
\label{eq:decoup}
H_f & = & \sum\nolimits_i t_i (f^\dagger_i f_{i+1} +\text{h.c.}) + \mu_i f^\dagger_i f_i \nonumber \\
H_\tau & = & \sum\nolimits_i h_i \tau^z_i + J_i \tau^x_i \tau^x_{i+1},
\ee
and each can be separately solved. The $f$ fermions form an Anderson insulator regardless of the disorder strength (since we are in one dimension). Meanwhile the $\tau$ spins are described by a RTFIM whose eigenstates can be either in a magnetically ordered (``paired'' or ``spin glass'') phase~\cite{HuseMBLQuantumOrder,KjallIsing,PekkerRSRGX}, when $h \ll J$, or a magnetically disordered (``paramagnetic'') phase, when $h \gg J$. (Here $h, J$ characterize the strength of randomness in $h_i, J_i$.) We argue that these limits correspond to the conventional and orthogonal Anderson-localized phases, respectively, and on perturbing away from exact solvability, lead to cMBL and oMBL phases. 
We note that, as outlined in~\cite{OrthogonalMetals}, the complex $f$ fermions carry \emph{all} the electric charge; thus, 
the transport properties are independent of the phase of the $\tau$ spins. In the $T\rightarrow\infty$ limit that is our main focus, all eigenstates are equiprobable and hence the thermodynamics in the two phases is identical (and trivial). 

Note that, although we primarily work with the solvable model and its free-fermion decoupling (\ref{eq:decoup}) for simplicity, the distinction between the orthogonal and conventional Anderson insulators is robust to perturbations of the solvable point that  lead to short-range interactions in the decoupled theories. Such interactions do not destroy the localized phase of the $f$ fermions nor disrupt the phase structure of the RTFIM. {Hence, we expect the eigenstates of the cMBL and oMBL phases to remain sharply distinct even in with such perturbations; the two phases must, therefore, be separated either by an eigenstate phase transition or an intermediate thermal phase.}

{\it Spectral Functions.---}In the clean case, a sharp diagnostic of the orthogonal metal is the behavior of its single-particle spectral function. As discussed in the introduction, the conventional metal has sharp peaks corresponding to quasiparticle excitations, whereas the orthogonal metal has an incoherent two-particle continuum, reflecting the fractionalization of the $c$ fermions. This diagnostic evidently fails in the localized phase, because \emph{any} local spectral function is dominated by a finite number (set by the localization length $\xi$) of sharp peaks. 

A second possibility is that the \emph{spatially averaged} spectral function retains a distinction between the two phases; as we shall see, this is true in the disordered system at $T = 0$ but not at $T > 0$. 
 Recall that the spectral function of an operator $\mathcal{O}$ is given by $A_{\mathcal{O}}(\omega) = -\frac{1}{\pi} \text{Im}\,G^{\text{ret}}_{\mathcal{O}}(\omega)$, where $G^{\text{ret}}_{\mathcal{O}}(t) \equiv -i \Theta(t) \text{Tr}\, \rho(\beta) \left[\mathcal{O}(t) ,\mathcal{O}^\dagger(0) \right]_\pm$ is the retarded Green's function of $\mathcal{O}$ and we choose a commutator (anticommutator) if $\mathcal{O}$ is bosonic (fermionic).  The use of a density matrix $\rho$ in this expression deserves comment in the context of MBL, where systems are usually considered in isolation. We imagine first coupling the system to a bath at temperature $T$ with strength $g$, and taking $g\rightarrow 0$ slowly so that the system remains in equilibrium with the bath~\cite{MBLBath}; then we may write $\rho = \mathcal{Z}^{-1} e^{-H/T}$, where $\mathcal{Z}$ is the partition function. In general, we have~\cite{SupMat} 
\be\label{eq:specfunc}
A_{\mathcal{O}}(\omega)  = \frac{1 \pm e^{-\beta \omega}}{\mathcal{Z}} \sum_{m, n} \left|\bra{m} \mathcal{O} \ket{n}\right|^2 e^{-\beta E_m} \delta(\omega - E_{mn})\,\,\,\,\,\,
\ee
where $\beta = 1/T$ is the inverse temperature, $m,n$ label many-body eigenstates, $E_{mn} \equiv E_n -E_m$, and we take the negative (positive) sign for bosonic (fermionic) $\mathcal{O}$. At the solvable point, we may express eigenstates of $H_{es}$ as tensor products of eigenstates of $H_\tau, H_f$, and decompose the energies, viz.
$\ket{m}  = \ket{m_\tau}\otimes \ket{m_f}$,  $E_{m} = E^\tau_{m_\tau} + E^{f}_{m_f}$.
Using this, the $c$ fermion spectral function is
\be\label{eq:cspecdec}
A_{c_j}(\omega)  &=&\frac{1 + e^{-\beta \omega}}{\mathcal{Z}} \!\!\sum_{\substack{m_\tau, m_f,\\ n_\tau, n_f}} \!\!\!\!\left|\bra{m_f}f_j \ket{n_f}\right|^2 e^{-\beta E^f_{m_f}} \\ & &\!\!\!\!\!\! \times \left|\bra{m_\tau}\tau^x_j \ket{n_\tau}\right|^2 e^{-\beta E^\tau_{m_\tau} } \delta(\omega -E^f_{m_f n_f} - E^\tau_{m_\tau n_\tau})\nonumber.
\ee
This can also be derived by expressing  $A_c$ in terms $f$s and $\tau$  spectral functions~\cite{SupMat}. {Although the factorization~\eqref{eq:cspecdec} is specific to the solvable point, our results are robust to small perturbations, as we discuss below.}

\begin{figure}[tt]
 \includegraphics[width=\columnwidth]{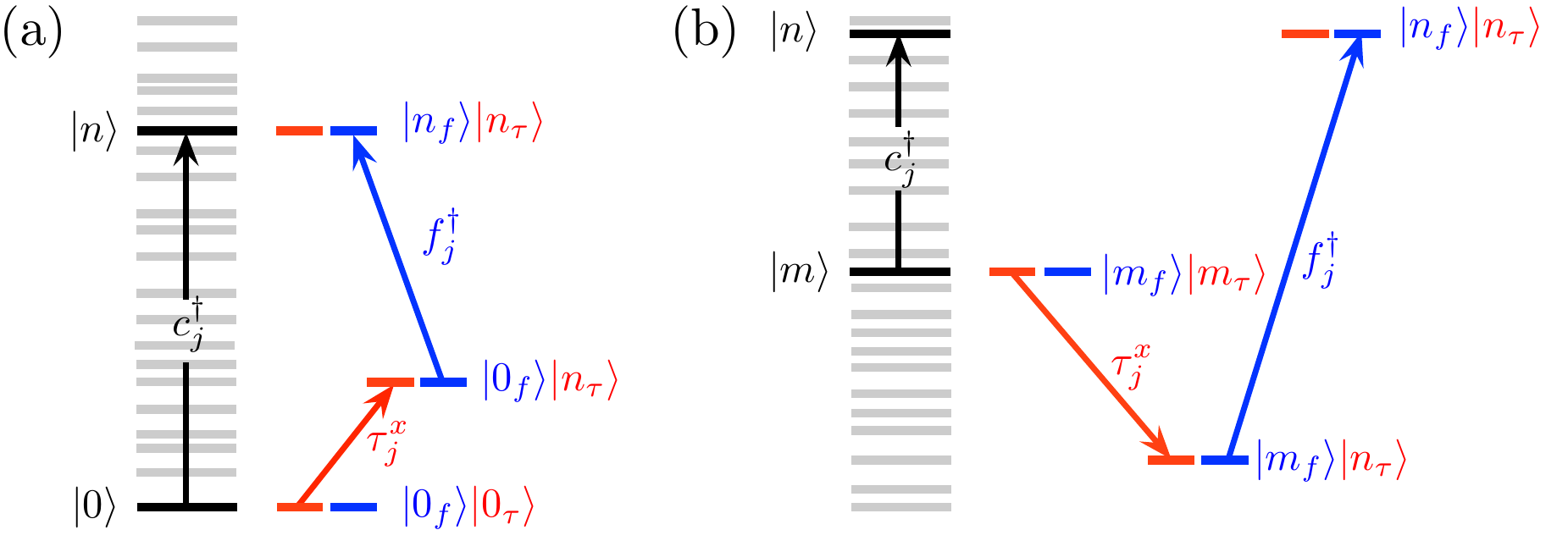}
 \caption{\label{fig:matrixelements}Processes contributing to  $A_c(\omega)$. (a) At $T=0$, only excitations about the ground state of $f$ and $\tau$ sectors contribute, restricting the phase space as each has $E>0$. (b) As $T\rightarrow\infty$,  arbitrary low-energy $c$-tunneling processes can be put `on shell' by trading energy between  sectors.}
  \end{figure}

 {\it Ground-State Spectral Functions.---} 
 The $T=0$ spectral function only involves excitations about the ground state in each sector; as these must have positive energy, for $\omega\rightarrow0$ only a limited set of transitions from each sector contribute to the spectral response (Fig.~\ref{fig:matrixelements}(a)). In this limit, is convenient to rewrite \eqref{eq:cspecdec} as a convolution of $T=0$ spectral functions of $f$ and $\tau$~\cite{SupMat},
  \be\label{eq:AcTeq0conv}
 A^0_{c_j}(\omega) &=& \int_0^\omega\,d\Omega A^0_{f_j}(\Omega) A^0_{\tau^x_j}(\omega - \Omega),
 \ee
 where  $A^0_{f_j}(\omega)  =\sum_{n_f} \left|\bra{0_f}f_j\ket{n_f}\right|^2 \delta(\omega - E^f_{n_f})$, $A^0_{\tau^x_j}(\omega) =\sum_{n_\tau}\left|\bra{0_\tau}\tau^x_j \ket{n_\tau}\right|^2 \delta(\omega- E^\tau_{n_\tau})$.

Let us consider the $f$ spectral function first, as its behavior is the same in both phases.
Since the initial and final states $\ket{0_f}$, $\ket{n_f}$ lie in distinct fermion-parity sectors, they do not experience mutual level repulsion, and so the energy $E^f_{n_f}$ can be arbitrarily small {(i.e., the tunneling density of states of an Anderson insulator is smooth and nonzero at the Fermi energy)}. 
Therefore, $\overline{A^0_{f_j} (\omega)} \overset{\omega\rightarrow0}{\approx} \nu_f >0$,  a constant, where the bar denotes both spatial and disorder-averaging.

We next consider the $\tau$ sector. In the spin glass phase, $\tau^x_j$ has a non-vanishing ground-state expectation value, $\bra{0_\tau} \tau^x_j \ket{0_\tau} =  m^x_j \neq 0$ leading to an $\omega=0$ contribution to the $\tau$ spectral function. This vanishes in the paramagnetic phase. {In addition, there are off-diagonal (i.e., finite-frequency) contributions to the $\tau$ spectral function. These go as continuously varying power laws~\cite{FisherRSRG1,FisherRSRG2}, $\overline{A_{\tau^x_j} (\omega)} \overset{\omega\rightarrow0}{\approx} m_x^2\delta(\omega) + K  (\omega/\omega_0)^{\gamma-1},$ where $m_x^2= \overline{(m^x_j)^2}$; $K$ is a constant; and $\gamma \geq 0$. In the exactly solvable model, $\gamma \rightarrow 0$ at the RTFIM transition. } Using \eqref{eq:AcTeq0conv}, 
\be
\overline{A^0_{c_j}(\omega)} \overset{\omega\rightarrow0}{\approx} \nu_f \left\{m_x^2+\nu_\tau  (\omega/\omega_0)^{\gamma}\right\}. 
\ee
 For $h\ll J$ the $\tau$s order ferromagnetically, $m_x\neq 0$ and there is a nonzero $\omega\rightarrow 0$ spectral response $\overline{A^0_{c_j}(\omega)} \sim \text{const.}$, corresponding to the conventional Anderson insulator. In the paramagnetic phase of the $\tau$s for $h\gg J$, $m_x =0$ and there is a `soft gap' to single-$c$-fermion tunneling: $\overline{A^0_{c_j}(\omega)} \sim \omega^{\gamma}$. This corresponds to the orthogonal Anderson insulator. The low-frequency $T=0$ spectral response is thus sharply distinct in the two phases.

  Note that we performed independent disorder averages for $f$ and $\tau$; this is no longer exact away from the solvable point. {Local interactions within  
  either sector are innocuous~\cite{PekkerRSRGX,QCGPRL}, as they will merely turn either noninteracting insulator into an MBL 
 insulator. 
Interactions will introduce correlations between the disorder experienced by the $f$ and $\tau$ sectors. The two distinct phases of the solvable model remain distinct in the presence of these disorder correlations, which preserve the Ising symmetry of the $\tau$ sector as well as the $U(1)$ symmetry of the charge sector.
Disorder correlations could in principle modify the low-frequency spectral behavior, but they are unlikely to when sufficiently weak, i.e., so long as the disorder \emph{induced} by sectors on each other remains small relative to the intrinsic disorder experienced by either.
}

 {\it Spectral Functions for $T > 0$.---} Crucial to the  $T=0$ spectral distinction between the cMBL and oMBL phases is that in order to tunnel a $c$ fermion, we must simultaneously tunnel an $f$ fermion and create a spin excitation, {both of which require a positive energy in the ground state.} This, coupled with the fact that the off-diagonal spin response vanishes as $\omega\rightarrow 0$, generates a soft tunneling gap in the oMBL phase. The situation is very different for $T > 0$. Here, the $c$ spectral function receives contributions from {\it all} initial states rather than just the ground state; therefore, a low-energy process for the $c$ fermions can built by offsetting a large positive energy difference in the $f$ sector by a large negative energy difference in the $\tau$ sector, or vice versa (Fig.~\ref{fig:matrixelements}(b))~\cite{mbmott}. In other words, at finite energy density the  low-frequency spectral response of the $c$s is generically built from  {high frequency} responses in the $f$ and $\tau$ sectors, and therefore the phase-space restrictions that lead to the distinct tunneling behavior in the oMBL and cMBL phases  are absent. Formally, this can be seen directly from the $T\rightarrow\infty$ limit of \eqref{eq:cspecdec}: all the Boltzmann factors equal the identity, and the sums range over all states in each sector, {leading to  indistinguishable spectral functions in the two phases.}

{A sharp spectral distinction is absent not just at $T = \infty$ but generically at all $T > 0$.}
However, for temperatures much lower than the characteristic {$\tau$ sector} energy scales,  
a crossover persists in $A_c(\omega)$. When the {$\tau$s are paramagnetic}, 
 the $c$ spectral function remains depleted at low frequencies, since only excitations with energies less than $T$ are appreciably populated in equilibrium and can thus undergo ``downhill'' transitions. We therefore expect that, in the {oMBL phase},
 $\overline{A_c(T, \omega \rightarrow 0)} \sim T^\gamma$.

 {\it Diagnostic at $T=\infty$.---} How can we distinguish between the orthogonal and conventional MBL phases for $T\rightarrow\infty$? 
{Recall that we are ultimately interested in systems where the $\tau$ spins are not physically observable, microscopic degrees of freedom to which one can couple directly; we must therefore build our desired diagnostic entirely out of $c$ fermions. Although the spatial correlations of $\tau^x$ exhibit spin glass order in one phase and decay exponentially in the other, one can only measure \emph{products} of $\tau$ and $f$ correlators. The $f$ correlators decay exponentially in both phases, since the $f$ fermions are always localized; therefore the $c$ correlators also decay exponentially in both phases. This overall exponential suppression masks the change from long-range behavior to exponential decay in the $\tau$ sector. We therefore seek a diagnostic sensitive to this qualitative change.}
  
\begin{figure}
\begin{center}
\includegraphics[width=\columnwidth]{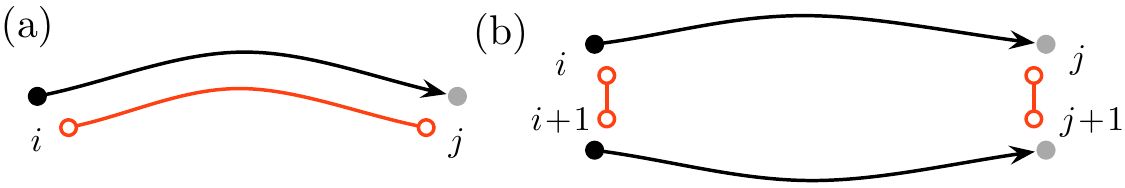}
\caption{Operator pairings  in Eq.~\eqref{eq:diagnostic}. Here, $f^\dagger$,$f$ are denoted by black (grey) filled circles, and $\tau^x$ by a red unfilled circle. The numerator involves pairing (a), whereas the denominator is dominated (in the $\tau$-paramagnet) by pairing  (b). In the $\tau$ spin glass, $\tau^x$ has a classical expectation value (the spin-glass order parameter) so at leading order  $\langle\tau^x_i \tau^x_j\rangle$ factorizes. \label{fdiags}}
\end{center}
\end{figure}

 The resolution is to examine the behavior of the following \emph{ratio} of correlators, {evaluated for simplicity at $T = \infty$ (i.e., summing over all eigenstates $m$ with equal weight)}:
 \be\label{eq:diagnostic}
 \mathcal{Q}(r)  =  \frac{ \overline{ \left({\sum_{i, m} \left|\bra{m} c^\dagger_i c_{i+r} \ket{m}\right|^2} \right)^2}}{\overline{\sum_{i,m} \left| \bra{m} c^\dagger_i c^\dagger_{i+1} c_{i+r} c_{i+r+1} \ket{m} \right|^2}}.
  \ee
The numerator of this expression is sensitive to the localization properties of both the $f$ and $\tau$ sectors; in the spin-glass phase it decays with the $f$-sector localization length (since the $\tau$ spins have spin-glass order), whereas in the paramagnet it decays with a localization length $\xi_c \simeq (1/\xi_f + 1/\xi_\tau)^{-1}$. The denominator, on the other hand, decays with the $f$-sector localization length in both phases. This is because, in the paramagnet, one can pair up the $\tau$ spins (which one can think of, via a Jordan-Wigner transformation, as real fermions) locally, but owing to charge conservation the complex $f$ fermions can only be paired up as in Fig.~\ref{fdiags}(b). 

Thus, the diagnostic~\eqref{eq:diagnostic} asymptotes to a constant value as $r \rightarrow \infty$ in the spin-glass phase of the $\tau$s, but decays exponentially in the paramagnet.
$\mathcal{Q}$ is experimentally accessible: its numerator is the $\omega\rightarrow 0$ limit of the spectral function of the operator $c^\dagger_i c_j$ that captures two-point correlations in tunneling, whereas the denominator is {again related to the zero-frequency limit of the spectral function of the pairing {operator $\mathcal{O}_j=c_j c_{j+1}$}.  In the {oMBL phase at $T=\infty$} 
fermionic pairs are more weakly localized than single fermions; this is a high-temperature manifestation of the relation between pairing and the orthogonal metal noted in
~\cite{OrthogonalMetals}. This method of ``factoring out'' a piece of a correlator to tease out the asymptotic behavior of one portion of it is reminiscent of the diagnostics in \cite{DiagnosingDeconfinement}, although the physics here is distinct.

{Three comments are in order. First, because squaring \emph{precedes} averaging over eigenstates, neither the numerator nor the denominator of $\mathcal{Q}(r)$ can be regarded as an equal-time correlator. Rather, they are $\omega \rightarrow 0$ limits of {(two-point)} spectral functions. Equal-time correlators depend only on the density matrix, whereas the $\omega \rightarrow 0$ limits of spectral functions are highly sensitive to the dynamics. Second, for imperfectly isolated systems~\cite{MBLBath}, $\mathcal{Q}(r)$ will behave thermally. However, {at finite frequencies the same ratio of spectral functions will continue to be small (large) in the oMBL (cMBL) phase.} 
Third, 
~\eqref{eq:diagnostic}, although introduced in the exactly solvable model, is robust {to interactions. Both} 
 the left and the right side of Fig.~\ref{fdiags} will  {then} involve terms with many $c$ and $c^\dagger$ operators. However, the number of long-distance pairings is determined by the \emph{imbalance} between $c^\dagger$s and $c$s, which does not change {as long as $U(1)$ symmetry is preserved.} 
 
 {\it Concluding remarks.---}
In closing, we discuss possible scenarios for the transition between 
the oMBL and cMBL phases. In the exactly solvable 1D model, this transition is accessed by tuning  $\delta = \text{var}h/\text{var}J$ and is controlled by an infinite-randomness critical point (IRCP), for both the ground~\cite{PhysRevLett.43.1434,PhysRevB.22.1305,FisherRSRG1,FisherRSRG2} and excited~\cite{HuseMBLQuantumOrder, PekkerRSRGX, QCGPRL} states. For the ground state, 
{the IRCP expected to be robust for generic interactions.
For excited states, it is at present unclear whether the IRCP  
survives away from the exactly solvable limit, or is replaced instead by a sliver of thermal phase. We expect that for weak interactions and not too near criticality, IRCP scaling will apply to our diagnostic.}
{We defer this, the possibility of universal finite-$T$ crossovers%
~\cite{KangPotterVasseurCrossover}, the critical behavior in $d>1$, and extensions to other fractionalized phases, to future work.}

{\it Acknowledgements.---} We thank R.~Nandkishore, M.~Serbyn, S.L.~Sondhi and R.~Vasseur for helpful discussions, and E.~Altman, J.T.~Chalker and A.C.~Potter for discussions and for comments on the manuscript. We acknowledge the hospitality of Trinity College, Cambridge, where part of this work was completed. S.A.P. is supported by NSF Grant DMR-1455366 and a UC President's Research Catalyst Award CA-15-327861, and  S.G. by the Burke Institute at Caltech.

\begin{thebibliography}{11}
\bibitem{note1} Note that this definition is not the same as in some texts, e.g. Mahan; consequently, the Lehmann representation differs from those by a factor of $2\pi$ in the integration measure.
\end{thebibliography}

%

\clearpage

\onecolumngrid
\section*{SUPPLEMENTARY INFORMATION}
\begin{appendix}

\section{$c$-Spectral Function for the Solvable Model}
\subsection{General Formula for Spectral Functions}
We  define the spectral function via~\cite{note1}
\be
A_{\mathcal{O}}(\omega) = -\frac{1}{\pi} \text{Im}\,G^{\text{ret}}_{\mathcal{O}}(\omega)
\ee
in terms of the retarded Green's function, which in the time domain is given by
\be
G^{\text{ret}}_{\mathcal{O}}(t) \equiv -i \Theta(t) \text{Tr} \left\{\rho(\beta) \left[\mathcal{O}(t) ,\mathcal{O}^\dagger(0) \right]_\pm\right\}
\ee 
where as usual $\beta = 1/T$, $\mathcal{O}(t) = e^{i Ht} \mathcal{O} e^{-iHt}$, $\rho(\beta)  = \frac{1}{Z}e^{-\beta H}$ is the density matrix, and we choose the commutator (anticommutator) for bosons (fermions) respectively. Inserting resolutions of the identity over many body eigenstates $\ket{m},\ket{n}$, it is straightforward to perform the Fourier transform, and in the end we arrive at 
\be
A_{\mathcal{O}}(\omega)  = \mathcal{Z}^{-1} (1 \pm e^{-\beta \omega}) \sum_{m,n} \left|\bra{m} \mathcal{O} \ket{n}\right|^2 e^{-\beta E_m} \delta(\omega - E_{mn})
\ee
where $E_{mn} =  E_n - E_m$ and we take the positive (negative) sign for fermions (bosons), and $\mathcal{Z}$ is the partition function.
\subsection{Direct Calculation}
Using the result of the preceding section, we have
\be
A_{c}(\omega)  = \mathcal{Z}^{-1} (1 + e^{-\beta \omega}) \sum_{m,n} \left|\bra{m}c \ket{n}\right|^2 e^{-\beta E_m} \delta(\omega - E_{mn})
\ee
At the solvable point, we may decompose every eigenstate of $H$ as a tensor product of eigenstates of $H_\tau, H_f$, via $\ket{m} = \ket{m_\tau} \ket{m_f}$,  with corresponding energy $E_m = E^\tau_{m_\tau} + E^f_{m_f}$. Using this, it is straightforward to write
\be\label{eq:direct}
A_{c}(\omega)  =  \mathcal{Z}^{-1}(1 + e^{-\beta \omega}) \sum_{\substack{m_\tau, m_f,\\ n_\tau, n_f}} \left|\bra{m_f}f \ket{n_f}\right|^2 \left|\bra{m_\tau}\tau^x \ket{n_\tau}\right|^2 e^{-\beta (E^\tau_{m_\tau} + E^f_{m_f})} \delta(\omega - E^f_{m_f n_f} - E^{\tau}_{m_\tau n_\tau}).
\ee

\subsection{Green's Function Calculation}
We now reproduce the result \eqref{eq:direct} in the Green's function formalism. Working in imaginary time at finite temperature, the Matsubara Green's function for the $c$ fermions is a convolution of that of the $f$ fermions and the slave spins:
\be
\mathcal{G}_c(i\omega_n) = \frac{1}{\beta} \sum_{m} 
\mathcal{G}_\tau(i\omega_n - i\Omega_m) \mathcal{G}_f(i\Omega_m)
\ee
where $\omega_n \equiv \frac{(2n+1)\pi}{\beta}, \Omega_m  \equiv \frac{(2m+1)\pi}{\beta}$ are fermionic Matsubara frequencies.
We now rewrite this using the Lehmann representation for the individual Green's functions: 
\be
\mathcal{G}_c(i\omega_n) =  \int {d\Omega_1} \int {d\Omega_2}\, A_\tau(\Omega_1) A_f(\Omega_2) \left[-\frac{1}{\beta} \sum_{m}  \frac{1}{i\Omega_m - (i\omega_n -\Omega_1)} \frac{1}{i\Omega_m - \Omega_2} \right].
\ee
We may perform the Matsubara sum by the usual contour integration trick:
\be
-\frac{1}{\beta} \sum_{m}  \frac{1}{i\Omega_m - (i\omega_n -\Omega_1)} \frac{1}{i\Omega_m - \Omega_2}
 =   \frac{ n_F(\Omega_2)- n_F(i\omega_n - \Omega_1)}{\Omega_1 + \Omega_2 - i\omega_n} =  \frac{ 1+ n_B(\Omega_1) - n_F(\Omega_2)}{ i\omega_n - (\Omega_1 + \Omega_2)}
\ee
where in the final step we used $n_F(i\omega_n - \Omega_1) = 1+ n_B(\Omega_1)$ for a fermionic Matsubara frequency $i\omega_n$. Thus, 
\be
\mathcal{G}_c(i\omega_n) =  \int {d\Omega_1} \int {d\Omega_2}\, A_\tau(\Omega_1) A_f(\Omega_2)   \frac{ 1+ n_B(\Omega_1) - n_F(\Omega_2)}{i\omega_n - (\Omega_1 + \Omega_2)}
\ee
Using  $A_{\mathcal{O}}(\omega) = -\frac{1}{\pi} \text{Im}\,G^{\text{ret}}_{\mathcal{O}}(\omega)$, with $G^{\text{ret}}_{\mathcal{O}}(\omega) = \mathcal{G}_{\mathcal{O}}(i\omega_n\rightarrow \omega + i0)$, and the Sokhotski-Plemelj identity $\text{Im}\frac{1}{x \pm i0} = \mp \pi  \delta(x) $ we find
\be
A_c (\omega) &=& \int {d\Omega_1} \int {d\Omega_2}\, A_\tau(\Omega_1) A_f(\Omega_2)   \left[{ 1+ n_B(\Omega_1) - n_F(\Omega_2)}\right]\delta(\omega -\Omega_1-\Omega_2)
=\int_{-\infty}^\infty d\Omega\, A_\tau(\omega - \Omega)A_f(\Omega) \mathcal{F}_T(\omega, \Omega)\nonumber\\
\ee
where  
\be
\mathcal{F}_T(\omega, \Omega) &=& 1+ n_B(\omega -\Omega) - n_F(\Omega) = (1+ n_B(\omega -\Omega))(1-n_F(\Omega)) (1 + e^{-\beta \omega})
\ee
is a temperature-dependent function of $\omega$ and $\Omega$. Therefore, we have
\be\label{eq:spectralfuncconvolution}
A_c (\omega) &=& (1 + e^{-\beta \omega})\int_{-\infty}^\infty d\Omega\, A_\tau(\omega - \Omega)A_f(\Omega)(1+ n_B(\omega -\Omega))(1-n_F(\Omega)).
\ee

It is straightforward to show that $\lim_{T\rightarrow0}\mathcal{F}_T(\omega, \Omega) = \Theta(\omega -\Omega) -\Theta(-\Omega)$, whence
\be
\lim_{T\rightarrow0} A_c(\omega) {=}\int_{0}^\omega d\Omega\, A_\tau(\omega - \Omega)A_f(\Omega).
\ee
thereby recovering the zero-temperature result quoted in \citesec{OrthoMetalsSup}.

We now compute the spectral functions of the $f$s and $\tau$s. Proceeding as before and using the slave spin decomposition, we have
\be
A_{f}(\omega)  &=&  \mathcal{Z}^{-1}(1 + e^{-\beta \omega}) \sum_{m,n} \left|\bra{m}f \ket{n}\right|^2 e^{-\beta E_m} \delta(\omega -  E_{mn})\nonumber\\
  &=& \mathcal{Z}^{-1}  (1 + e^{-\beta \omega}) \sum_{\substack{m_\tau, m_f,\\ n_\tau, n_f}} \left|\bra{m_f}f \ket{n_f}\right|^2 \left|\langle m_\tau |n_\tau \rangle\right|^2 e^{-\beta (E^\tau_{m_\tau} +E^f_{m_f})}\delta(\omega +E^\tau_{m_\tau} - E^f_{m_f n_f} - E^{\tau}_{m_\tau n_\tau})\nonumber\\
 &=& \mathcal{Z}^{-1} (1 + e^{-\beta \omega}) \sum_{m_\tau} e^{-\beta E^\tau_{m_\tau}} \sum_{m_f,n_f} \left|\bra{m_f}f \ket{n_f}\right|^2  e^{-\beta E^f_{m_f}} \delta(\omega - E^f_{m_f n_f} ) \nonumber\\
  &=&  (1 + e^{-\beta \omega}) \frac{\mathcal{Z}_\tau}{\mathcal{Z}}\sum_{m_f,n_f} \left|\bra{m_f}f \ket{n_f}\right|^2  e^{-\beta E^f_{m_f}} \delta(\omega - E^f_{m_f n_f} ),
\ee
where we have identified ${\mathcal{Z}_\tau}$, the slave spin partition function.

A similar calculation shows that
\be
A_{\tau}(\omega)  &=& (1 - e^{-\beta \omega}) \sum_{m,n} \left|\bra{m}\tau^x \ket{n}\right|^2 e^{-\beta E_m} \delta(\omega +E_m -E_n)\nonumber\\
 &=&(1 - e^{-\beta \omega})  \frac{\mathcal{Z}_f}{\mathcal{Z}}\sum_{m_\tau,n_\tau} \left|\bra{m_\tau}\tau^x \ket{n_\tau}\right|^2  e^{-\beta E^\tau_{m_\tau}} \delta(\omega - E^{\tau}_{m_\tau n_\tau}),
\ee
where ${\mathcal{Z}_f}$ is the partition function of the $f$ fermions.

We then have, using \eqref{eq:spectralfuncconvolution} and identifying ${\mathcal{Z}} ={\mathcal{Z}_f}{\mathcal{Z}_\tau}$, that the $c$ spectral function is given by
\be
A_{c}(\omega) &=& (1+ e^{-\beta \omega}) \int_{-\infty}^\infty d\Omega\, A_\tau(\omega - \Omega)A_f(\Omega) (1+ n_B(\omega -\Omega))(1-n_F(\Omega))
\nonumber\\
&=& \mathcal{Z}^{-1} (1 + e^{-\beta \omega})\sum_{\substack{m_\tau, m_f,\\ n_\tau, n_f}} \left|\bra{m_f}f \ket{n_f}\right|^2   \left|\bra{m_\tau}\tau^x \ket{n_\tau}\right|^2  e^{-\beta ( E^f_{m_f} + E^\tau_{m_\tau})}
\nonumber \\& & \times \int_{-\infty}^\infty d\Omega\, (1 - e^{-\beta (\omega-\Omega)}) (1 + e^{-\beta \Omega})  (1+ n_B(\omega -\Omega))(1-n_F(\Omega))
\delta(\Omega -  E^f_{m_f n_f}  ) \delta(\omega-\Omega - E^{\tau}_{m_\tau n_\tau} ). \nonumber\\
\ee
We now use the fact that $1 - e^{-\beta (\omega-\Omega)} =  (1+ n_B(\omega -\Omega))^{-1}$ and $1 + e^{-\beta \Omega} = (1-n_F(\Omega))^{-1}$ to simplify the integrand, resulting in
\be\label{eq:GreensFunc}
A_{c}(\omega)  &=& \mathcal{Z}^{-1} (1 + e^{-\beta \omega}) \sum_{\substack{m_\tau, m_f,\\ n_\tau, n_f}}\left|\bra{m_f}f \ket{n_f}\right|^2   \left|\bra{m_\tau}\tau^x \ket{n_\tau}\right|^2  e^{-\beta ( E^f_{m_f} + E^\tau_{m_\tau})}
\nonumber \\& &
 \times \int_{-\infty}^\infty \delta(\Omega -  E^f_{m_f n_f}  ) \delta(\omega-\Omega - E^{\tau}_{m_\tau n_\tau} ). \nonumber\\
 \nonumber
\\&=&  \mathcal{Z}^{-1} (1 + e^{-\beta \omega}) \sum_{\substack{m_\tau, m_f,\\ n_\tau, n_f}} \left|\bra{m_f}f \ket{n_f}\right|^2   \left|\bra{m_\tau}\tau^x \ket{n_\tau}\right|^2  e^{-\beta ( E^f_{m_f} + E^\tau_{m_\tau})}
\delta(\Omega -  E^f_{m_f n_f} - E^{\tau}_{m_\tau n_\tau} ). \nonumber\\
\ee
which is identical to the direct result \eqref{eq:direct}. Thus, the two approaches are exactly equivalent and yield identical results.

\section{Exactly Solvable Model and Duality in $d=2$}

In this section, we comment briefly on a dual description of the exactly solvable model in $d=2$. 
Let us consider the exactly solvable model introduced in the text, but in $d=2$:
\be\label{eq:HES2d}
H^{d=2}_{es} & = & \sum_{\langle i j \rangle} J_{ij} \sigma^x_i \sigma^x_{j} + \sum_i h_i \sigma^z_i (-1)^{c^\dagger_i c_i}  - \sum_{\langle i j\rangle} t_{ij} (c^\dagger_i \sigma^x_i \sigma^x_{j} c_{j} +\text{h.c.}) + \sum_i\mu_i c^\dagger_i c_i
\ee
As argued in the main text, we can reduce this to a decoupled model by defining $\tau^z_i \equiv \sigma^z_i (-1)^{f^\dagger_i f_i}$, $\tau^x_i = \sigma^x_i$, $f_i \equiv \tau^x_i c_i$: $H_{es} = H_f + H_\tau$, where
\be
H_f  =  \sum_{\langle i j\rangle } t_{ij} (f^\dagger_i f_{j} +\text{h.c.}) + \sum_i \mu_i f^\dagger_i f_i, \,\,\,\, H_\tau  =  \sum_i h_i \tau^z_i +  \sum_{\langle i j\rangle }J_{ij} \tau^x_i \tau^x_{j}.
\ee
This model is in the orthogonal Anderson insulator for $h\gg J$, and the conventional Anderson insulator in the opposite limit $J\gg g$. Assuming that MBL occurs in $d>2$ and following a line of reasoning similar to the main text, we conclude that this extends into a distinction between oMBL and cMBL phases away from the solvable point.

We now perform a duality transformation on the model \eqref{eq:HES2d}. Defining  link variables $\tilde{\sigma}^{\mu}_{ij}$ via $\sigma^x_i \sigma^x_{j} \equiv \tilde{\sigma}^x_{ij} $, 
$\sigma^z_{i} = \prod_{j \in +_i} \tilde{\sigma}^z_{ij}$ where the product is over all nearest neighbors of $i$ (i.e. $\prod_{j\in+_i}$ ranges over links on the `star' of $i$), we have
\be\label{eq:HES2dDUAL}
H^{d=2}_{es} & = & \sum_{\langle i j \rangle} J_{ij}\tilde{\sigma}^x_{ij}- \sum_i h_i (-1)^{c^\dagger_ic_i}\prod_{j \in +_i} \tilde{\sigma}^z_{ij}   - \sum_{\langle i j\rangle} t_{ij} (c^\dagger_i \tilde{\sigma}^x_{ij} c_{j} +\text{h.c.}) + \sum_i\mu_i c^\dagger_i c_i
\ee
However, an additional constraint emerges: if we consider the set of links on a plaquette $p$ of the direct lattice, their product is fixed:
\be
\prod_{\langle ij\rangle\in \Box_p} \tilde{\sigma}^x_{ij} = 1
\ee
as may be verified by explicit computation. This is required because we introduced twice as many $\tilde{\sigma}$ variables as there are $\sigma$ variables. Note that we have departed slightly from the convention of~\cite{OrthogonalMetals} where the duality transformation maps $\sigma^x$ to a function of $\tilde{\sigma}^z$s and vice versa; this is so that the resulting gauge theory is recast in a form amenable for comparison with other references.

As the lattice labeling above is potentially confusing, we rewrite it in more standard form before proceeding. In writing the dual theory (\ref{eq:HES2dDUAL})  we have continued to label sites, links and plaquettes in terms of the original, {\it direct} lattice. Recall that links of the direct lattice can be placed in one-one correspondence with links of the dual lattice; however, the plaquettes of the direct lattice map to `stars' of the dual lattice and vice versa (Fig.~). So, if we now rewrite the above expression in notation where $s$, $l$, and $p$,  label sites, links, and plaquettes of the dual lattice, we have
\be\label{eq:H2ESdualfinal}
H^{d=2}_{es} & = & \sum_{ l } J_{l}\tilde{\sigma}^x_{l}- \sum_p h_p (-1)^{c^\dagger_pc_p}\prod_{l \in \partial p } \tilde{\sigma}^z_{l}   - \sum_{ l; p_{1,2} \in p_l} t_{l } (c^\dagger_{p_1} \tilde{\sigma}^x_{l} c_{p_2} +\text{h.c.}) + \sum_p\mu_p c^\dagger_p c_p
\ee
operating on the Hilbert space of the $\tilde{\sigma}$ spins with the constraint
\be\label{eq:H2ESdualconstraint}
G_s = \prod_{ l :s \in \partial l} \tilde{\sigma}^x_{l} = 1
\ee
on all physical states.

Note that the fermionic variables now reside on plaquettes; we have used $\partial l$ to denote the sites at the ends of link $l$, and $p_l$ to denote plaquettes that share a link. This theory has a gauge redundancy: consider the  transformation 
\be
\tilde{\sigma}^{x}_{l} \rightarrow \left.\epsilon_{s_1} \tilde{\sigma}^{x}_{l} \epsilon_{s_2} \right|_{s_{1,2} = \partial l}
\ee
with $\epsilon_i = \pm 1$. This leaves the Hamiltonian invariant, and is generated by the operator $G_s$. 
Despite the tantalizing similarity of the expression above to a gauge theory minimally coupled to fermionic matter the $c$s  do not carry any $Z_2$ charge.; crucially, one notes that the hopping term commutes with the constraint and therefore moving a $c$ fermion does not require the addition of electric fluxes.  

Thus, we see that the model \eqref{eq:H2ESdualfinal} supplemented with constraint \eqref{eq:H2ESdualconstraint} (which could also be imposed as a `soft', energetic constraint) is one where the cMBL-oMBL transition can occur without a global order parameter. In $d=1$ duality maps the Ising model back into itself, exchanging ordered and disordered phases. There, performing the duality leads to a model where the cMBL phase corresponds to the paramagnetic phase and the oMBL phase to the spin glass phase, but there remains a local order parameter in the spin sector.
\end{appendix}


\end{document}